\documentclass[letterpaper,compsoc,twoside]{IEEEtran}
\usepackage{fixltx2e} 
\usepackage{cmap} 
\usepackage{ifthen}
\usepackage[T1]{fontenc}
\usepackage[utf8]{inputenc}
\usepackage{amsmath}

\usepackage[font={small,it},labelfont=bf]{caption}
\usepackage{float}

\usepackage{longtable,ltcaption,array}
\newlength{\DUtablewidth} 

\pdfoutput=1
\usepackage{scipy}
\makeatletter
\def\PY@reset{\let\PY@it=\relax \let\PY@bf=\relax%
    \let\PY@ul=\relax \let\PY@tc=\relax%
    \let\PY@bc=\relax \let\PY@ff=\relax}
\def\PY@tok#1{\csname PY@tok@#1\endcsname}
\def\PY@toks#1+{\ifx\relax#1\empty\else%
    \PY@tok{#1}\expandafter\PY@toks\fi}
\def\PY@do#1{\PY@bc{\PY@tc{\PY@ul{%
    \PY@it{\PY@bf{\PY@ff{#1}}}}}}}
\def\PY#1#2{\PY@reset\PY@toks#1+\relax+\PY@do{#2}}

\expandafter\def\csname PY@tok@gd\endcsname{\def\PY@tc##1{\textcolor[rgb]{0.63,0.00,0.00}{##1}}}
\expandafter\def\csname PY@tok@gu\endcsname{\let\PY@bf=\textbf\def\PY@tc##1{\textcolor[rgb]{0.50,0.00,0.50}{##1}}}
\expandafter\def\csname PY@tok@gt\endcsname{\def\PY@tc##1{\textcolor[rgb]{0.00,0.27,0.87}{##1}}}
\expandafter\def\csname PY@tok@gs\endcsname{\let\PY@bf=\textbf}
\expandafter\def\csname PY@tok@gr\endcsname{\def\PY@tc##1{\textcolor[rgb]{1.00,0.00,0.00}{##1}}}
\expandafter\def\csname PY@tok@cm\endcsname{\let\PY@it=\textit\def\PY@tc##1{\textcolor[rgb]{0.25,0.50,0.56}{##1}}}
\expandafter\def\csname PY@tok@vg\endcsname{\def\PY@tc##1{\textcolor[rgb]{0.73,0.38,0.84}{##1}}}
\expandafter\def\csname PY@tok@m\endcsname{\def\PY@tc##1{\textcolor[rgb]{0.13,0.50,0.31}{##1}}}
\expandafter\def\csname PY@tok@mh\endcsname{\def\PY@tc##1{\textcolor[rgb]{0.13,0.50,0.31}{##1}}}
\expandafter\def\csname PY@tok@cs\endcsname{\def\PY@tc##1{\textcolor[rgb]{0.25,0.50,0.56}{##1}}\def\PY@bc##1{\setlength{\fboxsep}{0pt}\colorbox[rgb]{1.00,0.94,0.94}{\strut ##1}}}
\expandafter\def\csname PY@tok@ge\endcsname{\let\PY@it=\textit}
\expandafter\def\csname PY@tok@vc\endcsname{\def\PY@tc##1{\textcolor[rgb]{0.73,0.38,0.84}{##1}}}
\expandafter\def\csname PY@tok@il\endcsname{\def\PY@tc##1{\textcolor[rgb]{0.13,0.50,0.31}{##1}}}
\expandafter\def\csname PY@tok@go\endcsname{\def\PY@tc##1{\textcolor[rgb]{0.20,0.20,0.20}{##1}}}
\expandafter\def\csname PY@tok@cp\endcsname{\def\PY@tc##1{\textcolor[rgb]{0.00,0.44,0.13}{##1}}}
\expandafter\def\csname PY@tok@gi\endcsname{\def\PY@tc##1{\textcolor[rgb]{0.00,0.63,0.00}{##1}}}
\expandafter\def\csname PY@tok@gh\endcsname{\let\PY@bf=\textbf\def\PY@tc##1{\textcolor[rgb]{0.00,0.00,0.50}{##1}}}
\expandafter\def\csname PY@tok@ni\endcsname{\let\PY@bf=\textbf\def\PY@tc##1{\textcolor[rgb]{0.84,0.33,0.22}{##1}}}
\expandafter\def\csname PY@tok@nl\endcsname{\let\PY@bf=\textbf\def\PY@tc##1{\textcolor[rgb]{0.00,0.13,0.44}{##1}}}
\expandafter\def\csname PY@tok@nn\endcsname{\let\PY@bf=\textbf\def\PY@tc##1{\textcolor[rgb]{0.05,0.52,0.71}{##1}}}
\expandafter\def\csname PY@tok@no\endcsname{\def\PY@tc##1{\textcolor[rgb]{0.38,0.68,0.84}{##1}}}
\expandafter\def\csname PY@tok@na\endcsname{\def\PY@tc##1{\textcolor[rgb]{0.25,0.44,0.63}{##1}}}
\expandafter\def\csname PY@tok@nb\endcsname{\def\PY@tc##1{\textcolor[rgb]{0.00,0.44,0.13}{##1}}}
\expandafter\def\csname PY@tok@nc\endcsname{\let\PY@bf=\textbf\def\PY@tc##1{\textcolor[rgb]{0.05,0.52,0.71}{##1}}}
\expandafter\def\csname PY@tok@nd\endcsname{\let\PY@bf=\textbf\def\PY@tc##1{\textcolor[rgb]{0.33,0.33,0.33}{##1}}}
\expandafter\def\csname PY@tok@ne\endcsname{\def\PY@tc##1{\textcolor[rgb]{0.00,0.44,0.13}{##1}}}
\expandafter\def\csname PY@tok@nf\endcsname{\def\PY@tc##1{\textcolor[rgb]{0.02,0.16,0.49}{##1}}}
\expandafter\def\csname PY@tok@si\endcsname{\let\PY@it=\textit\def\PY@tc##1{\textcolor[rgb]{0.44,0.63,0.82}{##1}}}
\expandafter\def\csname PY@tok@s2\endcsname{\def\PY@tc##1{\textcolor[rgb]{0.25,0.44,0.63}{##1}}}
\expandafter\def\csname PY@tok@vi\endcsname{\def\PY@tc##1{\textcolor[rgb]{0.73,0.38,0.84}{##1}}}
\expandafter\def\csname PY@tok@nt\endcsname{\let\PY@bf=\textbf\def\PY@tc##1{\textcolor[rgb]{0.02,0.16,0.45}{##1}}}
\expandafter\def\csname PY@tok@nv\endcsname{\def\PY@tc##1{\textcolor[rgb]{0.73,0.38,0.84}{##1}}}
\expandafter\def\csname PY@tok@s1\endcsname{\def\PY@tc##1{\textcolor[rgb]{0.25,0.44,0.63}{##1}}}
\expandafter\def\csname PY@tok@gp\endcsname{\let\PY@bf=\textbf\def\PY@tc##1{\textcolor[rgb]{0.78,0.36,0.04}{##1}}}
\expandafter\def\csname PY@tok@sh\endcsname{\def\PY@tc##1{\textcolor[rgb]{0.25,0.44,0.63}{##1}}}
\expandafter\def\csname PY@tok@ow\endcsname{\let\PY@bf=\textbf\def\PY@tc##1{\textcolor[rgb]{0.00,0.44,0.13}{##1}}}
\expandafter\def\csname PY@tok@sx\endcsname{\def\PY@tc##1{\textcolor[rgb]{0.78,0.36,0.04}{##1}}}
\expandafter\def\csname PY@tok@bp\endcsname{\def\PY@tc##1{\textcolor[rgb]{0.00,0.44,0.13}{##1}}}
\expandafter\def\csname PY@tok@c1\endcsname{\let\PY@it=\textit\def\PY@tc##1{\textcolor[rgb]{0.25,0.50,0.56}{##1}}}
\expandafter\def\csname PY@tok@kc\endcsname{\let\PY@bf=\textbf\def\PY@tc##1{\textcolor[rgb]{0.00,0.44,0.13}{##1}}}
\expandafter\def\csname PY@tok@c\endcsname{\let\PY@it=\textit\def\PY@tc##1{\textcolor[rgb]{0.25,0.50,0.56}{##1}}}
\expandafter\def\csname PY@tok@mf\endcsname{\def\PY@tc##1{\textcolor[rgb]{0.13,0.50,0.31}{##1}}}
\expandafter\def\csname PY@tok@err\endcsname{\def\PY@bc##1{\setlength{\fboxsep}{0pt}\fcolorbox[rgb]{1.00,0.00,0.00}{1,1,1}{\strut ##1}}}
\expandafter\def\csname PY@tok@kd\endcsname{\let\PY@bf=\textbf\def\PY@tc##1{\textcolor[rgb]{0.00,0.44,0.13}{##1}}}
\expandafter\def\csname PY@tok@ss\endcsname{\def\PY@tc##1{\textcolor[rgb]{0.32,0.47,0.09}{##1}}}
\expandafter\def\csname PY@tok@sr\endcsname{\def\PY@tc##1{\textcolor[rgb]{0.14,0.33,0.53}{##1}}}
\expandafter\def\csname PY@tok@mo\endcsname{\def\PY@tc##1{\textcolor[rgb]{0.13,0.50,0.31}{##1}}}
\expandafter\def\csname PY@tok@mi\endcsname{\def\PY@tc##1{\textcolor[rgb]{0.13,0.50,0.31}{##1}}}
\expandafter\def\csname PY@tok@kn\endcsname{\let\PY@bf=\textbf\def\PY@tc##1{\textcolor[rgb]{0.00,0.44,0.13}{##1}}}
\expandafter\def\csname PY@tok@o\endcsname{\def\PY@tc##1{\textcolor[rgb]{0.40,0.40,0.40}{##1}}}
\expandafter\def\csname PY@tok@kr\endcsname{\let\PY@bf=\textbf\def\PY@tc##1{\textcolor[rgb]{0.00,0.44,0.13}{##1}}}
\expandafter\def\csname PY@tok@s\endcsname{\def\PY@tc##1{\textcolor[rgb]{0.25,0.44,0.63}{##1}}}
\expandafter\def\csname PY@tok@kp\endcsname{\def\PY@tc##1{\textcolor[rgb]{0.00,0.44,0.13}{##1}}}
\expandafter\def\csname PY@tok@w\endcsname{\def\PY@tc##1{\textcolor[rgb]{0.73,0.73,0.73}{##1}}}
\expandafter\def\csname PY@tok@kt\endcsname{\def\PY@tc##1{\textcolor[rgb]{0.56,0.13,0.00}{##1}}}
\expandafter\def\csname PY@tok@sc\endcsname{\def\PY@tc##1{\textcolor[rgb]{0.25,0.44,0.63}{##1}}}
\expandafter\def\csname PY@tok@sb\endcsname{\def\PY@tc##1{\textcolor[rgb]{0.25,0.44,0.63}{##1}}}
\expandafter\def\csname PY@tok@k\endcsname{\let\PY@bf=\textbf\def\PY@tc##1{\textcolor[rgb]{0.00,0.44,0.13}{##1}}}
\expandafter\def\csname PY@tok@se\endcsname{\let\PY@bf=\textbf\def\PY@tc##1{\textcolor[rgb]{0.25,0.44,0.63}{##1}}}
\expandafter\def\csname PY@tok@sd\endcsname{\let\PY@it=\textit\def\PY@tc##1{\textcolor[rgb]{0.25,0.44,0.63}{##1}}}

\def\PYZbs{\char`\\}
\def\PYZus{\char`\_}

\def\PYZlt{\char`\<}

\def\PYZdq{\char`\"}

\makeatother

\providecommand*{\DUrole}[2]{%
  \ifcsname DUrole#1\endcsname%
    \csname DUrole#1\endcsname{#2}%
  \else
    \ifcsname docutilsrole#1\endcsname%
      \csname docutilsrole#1\endcsname{#2}%
    \else%
      #2%
    \fi%
  \fi%
}

\ifthenelse{\isundefined{\hypersetup}}{
  \usepackage[colorlinks=true,linkcolor=blue,urlcolor=blue]{hyperref}
  \urlstyle{same} 
}{}

\begin{document}
\newcounter{footnotecounter}\title{PyFAI: a Python library for high performance azimuthal integration on GPU}\author{Jérôme Kieffer$^{\setcounter{footnotecounter}{1}\fnsymbol{footnotecounter}\setcounter{footnotecounter}{2}\fnsymbol{footnotecounter}}$%
          \setcounter{footnotecounter}{1}\thanks{\fnsymbol{footnotecounter} %
          Corresponding author: \protect\href{mailto:jerome.kieffer@esrf.fr}{jerome.kieffer@esrf.fr}}\setcounter{footnotecounter}{2}\thanks{\fnsymbol{footnotecounter} European Synchrotron Radiation Facility, Grenoble, France}, Giannis Ashiotis$^{\setcounter{footnotecounter}{2}\fnsymbol{footnotecounter}}$\thanks{%

          \noindent%
          Copyright\,\copyright\,2014 Jérôme Kieffer et al. This is an open-access article distributed under the terms of the Creative Commons Attribution License, which permits unrestricted use, distribution, and reproduction in any medium, provided the original author and source are credited. http://creativecommons.org/licenses/by/3.0/%
        }}\maketitle
          \renewcommand{\leftmark}{PROC. OF THE 7th EUR. CONF. ON PYTHON IN SCIENCE (EUROSCIPY 2014)}
          \renewcommand{\rightmark}{PYFAI: A PYTHON LIBRARY FOR HIGH PERFORMANCE AZIMUTHAL INTEGRATION ON GPU}

\setcounter{page}{3}
\newcommand*{\docutilsroleref}{\ref}
\newcommand*{\docutilsrolelabel}{\label}
\AtEndDocument{\cleardoublepage}
\begin{abstract}The pyFAI package has been designed to reduce X-ray diffraction images
into powder diffraction curves to be further processed by scientists.
This contribution describes how to convert an image into a radial profile
using the Numpy package, how the process was accelerated using Cython.
The algorithm was parallelised, needing a complete re-design to benefit
from massively parallel devices like graphical processing units or accelerators like
the Intel Xeon Phi using the PyOpenCL library.\end{abstract}\begin{IEEEkeywords}X-rays, powder diffraction, SAXS, HPC, parallel algorithms, OpenCL.\end{IEEEkeywords}

\section{Introduction%
  \label{introduction}%
}

The Python programming language is widely adopted in the scientific community
and especially in crystallography, this is why a convenient azimuthal integration
routine, one of the basic algorithms in crystallography, was requested by the synchrotron community.
The advent of pixel-detectors with their very high speed (up to 3000 frames per second)
imposed strong constraints in speed that most available programs (\cite{FIT2D}, \cite{SPD}, ...),
written in Fortran or C, could not meet.

The \cite{pyFAI} project started in 2011 and aims at providing a convenient \emph{Pythonic} interface
for azimuthal integration, so that any diffractionist can adapt it to the type of experiment
he is interested in.
This contribution describes how one of the most fundamental
algorithms used in crystallography has been implemented in Python
and how it was accelerated to match the readout speeds of today's fastest detectors.

After describing a typical experiment and explaining what is measured and how it must be transformed (section 2),
section 3 describes how the algorithm can be vectorised using \cite{NumPy} and sped up with \cite{Cython}.
Section 4 highlights the accuracy enhancement recently introduced while section 5 focuses on
the parallelisation of the azimuthal integration task on many-core systems like Graphical Processing Units (GPU) or on accelerators via \cite{PyOpenCL}.
In section 6, serial and parallel implementations using \cite{OpenMP} and \cite{OpenCL} from various vendors and devices are benchmarked.

\section{Description of the experiment%
  \label{description-of-the-experiment}%
}

X-rays are electromagnetic waves, similar to visible light, except for their wavelengths which are much shorter,
typically of the size of inter-atomic distances, making them a perfect probe to analyse atomic and molecular structures.
X-rays can be elastically scattered (i.e. re-emitted with the same energy) by the electron cloud surrounding atoms.
When atoms are arranged periodically, as in a crystal, scattered X-rays interfere in a constructive way
when the difference of their optical paths is a multiple of the wavelength: $2d sin(\theta) = n\lambda$.
In this formula, known as \emph{Bragg's law}, \emph{d} is the distance between crystal plans, $\theta$ is the incidence angle, $\lambda$ is the wavelength and n is an integer.
An X-ray beam crossing a powder-like sample made of many randomly oriented small crystals is then scattered along multiple concentric cones.
In a powder diffraction experiment, one aims at measuring the intensity of X-rays as a function of the cone aperture, averaged along each ring.
This transformation is called \textquotedbl{}azimuthal integration\textquotedbl{} as it is an averaging of the signal along the azimuthal angle.\begin{figure}[]\noindent\makebox[\columnwidth][c]{\includegraphics[width=\columnwidth]{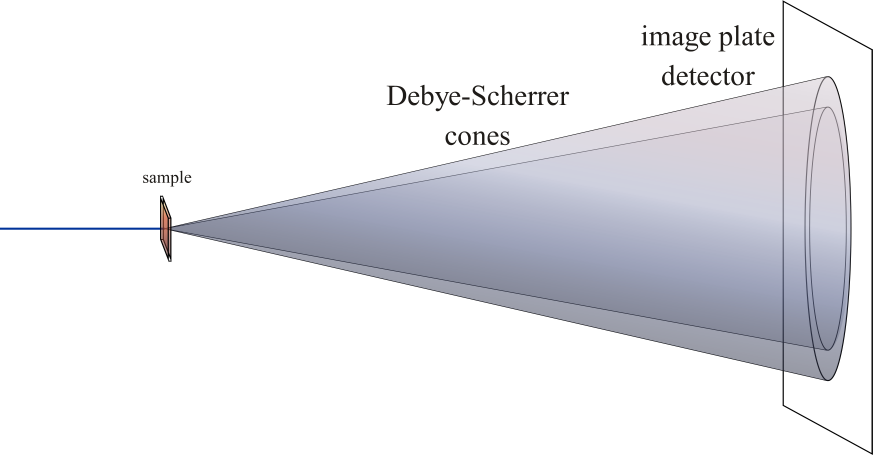}}
\caption{Debye-Scherrer cones obtained from diffraction of a monochromatic X-Ray beam by a powder of crystallised material. (Credits: CC-BY-SA  Klaus-Dieter Liss) \DUrole{label}{diffraction}}
\end{figure}

\section{Azimuthal integration%
  \label{azimuthal-integration}%
}

While pyFAI addresses the needs of both mono- and bi-dimensional integration in different spaces (real or reciprocal),
this contribution focuses on the algorithmic and implementation part of the method.
The coordinates in which pyFAI operates can be one of the following:%
\begin{itemize}

\item 

$r = \sqrt{x^2+y^2}$
\item 

$\chi = tan^{-1}(y/x)$
\item 

$2\theta = tan^{-1}(r/d)$
\item 

$q = 4 \pi sin({2 \theta} / 2)/ \lambda$
\end{itemize}

The pyFAI library was designed to feature a \emph{Pythonic} interface and work together with \cite{FabIO} for image reading (or \cite{H5Py} for HDF5 files).
The following snippet of code explains the basic usage of the library: \DUrole{label}{use}\begin{Verbatim}[commandchars=\\\{\},fontsize=\footnotesize]
\PY{k+kn}{import} \PY{n+nn}{fabio}\PY{o}{,} \PY{n+nn}{pyFAI}
\PY{n}{data} \PY{o}{=} \PY{n}{fabio}\PY{o}{.}\PY{n}{open}\PY{p}{(}\PY{l+s}{\PYZdq{}}\PY{l+s}{Pilatus1M.edf}\PY{l+s}{\PYZdq{}}\PY{p}{)}\PY{o}{.}\PY{n}{data}
\PY{n}{ai} \PY{o}{=} \PY{n}{pyFAI}\PY{o}{.}\PY{n}{load}\PY{p}{(}\PY{l+s}{\PYZdq{}}\PY{l+s}{Pilatus1M.poni}\PY{l+s}{\PYZdq{}}\PY{p}{)}
\PY{n}{tth}\PY{p}{,} \PY{n}{I} \PY{o}{=} \PY{n}{ai}\PY{o}{.}\PY{n}{integrate1d}\PY{p}{(}\PY{n}{data}\PY{p}{,} \PY{l+m+mi}{1000}\PY{p}{,} \PY{n}{unit}\PY{o}{=}\PY{l+s}{\PYZdq{}}\PY{l+s}{2th\PYZus{}deg}\PY{l+s}{\PYZdq{}}\PY{p}{,}\PYZbs{}
                                     \PY{n}{method}\PY{o}{=}\PY{l+s}{\PYZdq{}}\PY{l+s}{numpy}\PY{l+s}{\PYZdq{}}\PY{p}{)}
\end{Verbatim}
Output variables' space (\emph{r}, \emph{q} or $2\theta$) and units can be chosen with the \emph{unit} keyword.
The \emph{method} keyword is used to choose one of the available algorithms for the integration.
These algorithms will be described in this contribution.
However, the experiment reported here will be limited to 1D full azimuthal integration, with a planar detector, orthogonal to the incoming beam.
In this case the conics described by the beam on the detector are concentric circles.
The generic geometry used in pyFAI has already been described in \cite{pyFAI_ocl}.

\subsection{Test case%
  \label{test-case}%
}

To let the reader have an idea of the scale of the problem and the performances needed, we will work on
the simulated image of gold powder diffracting an X-ray beam of wavelength = 1.0e-10m (the intensity of all rings is the same).
The detector, which has a pixel size of 1e-4m (2048x2048 pixels), is placed at a distance of 0.1 m from the sample, orthogonal to the incident beam, which coincides with the centre of the rings.
Figure \DUrole{ref}{rings} represents the input diffraction image (upper part) and the integrated profile along the azimuthal angle (lower part).
The radial unit in this case is simply the radius calculated from $r=\sqrt{(x - x_c)^2 + (y - y_c)^2}$,
while crystallographers would have preferred $2\theta$ or the scattering vector length \emph{q}.\begin{figure}[]\noindent\makebox[\columnwidth][c]{\includegraphics[width=\columnwidth]{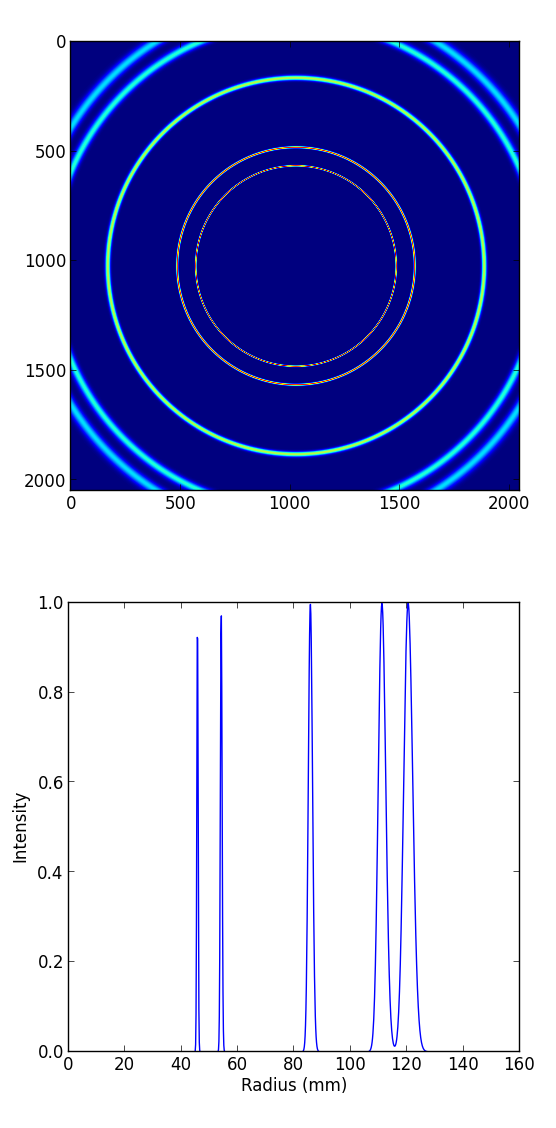}}
\caption{Simulated powder diffraction image (top) and integrated profile (bottom).  \DUrole{label}{rings}}
\end{figure}

\subsection{Naive implementation%
  \label{naive-implementation}%
}

The initial step of any implementation is calculating the radius array, from the previous formula.
Using Numpy's slicing feature one can extract all pixels which are between r1 and r2 and average out their values:\begin{Verbatim}[commandchars=\\\{\},fontsize=\footnotesize]
\PY{k}{def} \PY{n+nf}{azimint\PYZus{}naive}\PY{p}{(}\PY{n}{data}\PY{p}{,} \PY{n}{npt}\PY{p}{,} \PY{n}{radius}\PY{p}{)}\PY{p}{:}
    \PY{n}{rmax} \PY{o}{=} \PY{n}{radius}\PY{o}{.}\PY{n}{max}\PY{p}{(}\PY{p}{)}
    \PY{n}{res} \PY{o}{=} \PY{n}{numpy}\PY{o}{.}\PY{n}{zeros}\PY{p}{(}\PY{n}{npt}\PY{p}{)}
    \PY{k}{for} \PY{n}{i} \PY{o+ow}{in} \PY{n+nb}{range}\PY{p}{(}\PY{n}{npt}\PY{p}{)}\PY{p}{:}
        \PY{n}{r1} \PY{o}{=} \PY{n}{rmax} \PY{o}{*} \PY{n}{i} \PY{o}{/} \PY{n}{npt}
        \PY{n}{r2} \PY{o}{=} \PY{n}{rmax} \PY{o}{*} \PY{p}{(}\PY{n}{i}\PY{o}{+}\PY{l+m+mi}{1}\PY{p}{)} \PY{o}{/} \PY{n}{npt}
        \PY{n}{mask\PYZus{}r12} \PY{o}{=} \PY{n}{numpy}\PY{o}{.}\PY{n}{logical\PYZus{}and}\PY{p}{(}\PY{p}{(}\PY{n}{r1} \PY{o}{\PYZlt{}}\PY{o}{=} \PY{n}{radius}\PY{p}{)}\PY{p}{,}
                     \PY{p}{(}\PY{n}{radius} \PY{o}{\PYZlt{}} \PY{n}{r2}\PY{p}{)}\PY{p}{)}
        \PY{n}{values\PYZus{}r12} \PY{o}{=} \PY{n}{data}\PY{p}{[}\PY{n}{mask\PYZus{}r12}\PY{p}{]}
        \PY{n}{res}\PY{p}{[}\PY{n}{i}\PY{p}{]} \PY{o}{=} \PY{n}{values\PYZus{}r12}\PY{o}{.}\PY{n}{mean}\PY{p}{(}\PY{p}{)}
    \PY{k}{return} \PY{n}{res}
\end{Verbatim}
The slicing operation takes tens of milliseconds and needs to be repeated thousands of times for a single image,
making each integration last 40 seconds, something that is unacceptably slow. \DUrole{label}{naive}

\subsection{Numpy histograms%
  \label{numpy-histograms}%
}

The naive formulation in section \DUrole{ref}{naive} can be re-written using histograms.
The \emph{mean} call can be replaced with the ratio of the sum of all values divided by the number of contributing pixels:\begin{Verbatim}[commandchars=\\\{\},fontsize=\footnotesize]
\PY{n}{values\PYZus{}r12}\PY{o}{.}\PY{n}{mean}\PY{p}{(}\PY{p}{)} \PY{o}{=} \PY{n}{values\PYZus{}r12}\PY{o}{.}\PY{n}{sum}\PY{p}{(}\PY{p}{)} \PY{o}{/} \PY{n}{mask\PYZus{}r12}\PY{o}{.}\PY{n}{sum}\PY{p}{(}\PY{p}{)}
\end{Verbatim}
The denominator, \emph{mask\_r12.sum()}, can be obtained from the histogram of \emph{radius} values and the numerator, \emph{values\_r12.sum()}
from the histogram of \emph{radius} weighted by the \emph{data} in the image:\begin{Verbatim}[commandchars=\\\{\},fontsize=\footnotesize]
\PY{k}{def} \PY{n+nf}{azimint\PYZus{}hist}\PY{p}{(}\PY{n}{data}\PY{p}{,} \PY{n}{npt}\PY{p}{,} \PY{n}{radius}\PY{p}{)}\PY{p}{:}
    \PY{n}{histu} \PY{o}{=} \PY{n}{np}\PY{o}{.}\PY{n}{histogram}\PY{p}{(}\PY{n}{radius}\PY{p}{,} \PY{n}{npt}\PY{p}{)}\PY{p}{[}\PY{l+m+mi}{0}\PY{p}{]}
    \PY{n}{histw} \PY{o}{=} \PY{n}{np}\PY{o}{.}\PY{n}{histogram}\PY{p}{(}\PY{n}{radius}\PY{p}{,} \PY{n}{npt}\PY{p}{,} \PY{n}{weights}\PY{o}{=}\PY{n}{data}\PY{p}{)}\PY{p}{[}\PY{l+m+mi}{0}\PY{p}{]}
    \PY{k}{return} \PY{n}{histw} \PY{o}{/} \PY{n}{histu}
\end{Verbatim}
This implementation takes about 800ms which is much faster than the loop written in Python,
but can be optimised by reading the radius array from central memory only once.

\subsection{Cython implementation%
  \label{cython-implementation}%
}

Histograms were re-implemented using \cite{Cython} to generate simultaneously both the
weighted and the unweighted histograms with a single memory read of the radius array.
The better use of the CPU cache decreases the integration time significantly (down to 150ms on a single core).

\subsubsection{OpenMP support in Cython%
  \label{openmp-support-in-cython}%
}

To accelerate further the code we decided to parallelise the \cite{Cython} code using \cite{OpenMP}.
While the implementation was fast, the results we obtained were wrong (by a few percent) due to
write conflicts, not protected by atomic\_add operations.
Apparently the use of atomic operation is still not yet possible in \cite{Cython} (summer 2014).
Multi-threaded histogramming was made possible by having several threads running simultaneously, each working on a separate histogram,
which implies the allocation of much more memory for output arrays.\begin{table}
\setlength{\DUtablewidth}{0.8\linewidth}
\begin{longtable*}[c]{|p{0.203\DUtablewidth}|p{0.203\DUtablewidth}|}
\hline

Implement. & 

Exec. time (ms) \\
\hline

loop + mean & 

44000 \\
\hline

np.histogram & 

829 \\
\hline

Cython 1\_th & 

149 \\
\hline

Cython 2\_th & 

81 \\
\hline

Cython 4\_th & 

59 \\
\hline

Cython 8\_th & 

41 \\
\hline

Cython 16\_th & 

48 \\
\hline
\end{longtable*}
\caption{Azimuthal integration time for a 4 Mpix image measured on two Xeon E5520 (2x 4-core hyper-threaded at 2.2 GHz) \DUrole{label}{Cython}}\end{table}

The gains in performance obtained by this method (see table \DUrole{ref}{Cython}) were minor, especially when using more than 2 threads,
illustrating the limits of the paralellisation scheme.
The only way to go faster is to start thinking in parallel from the beginning
and re-design the algorithm so that it works natively with lots of threads.
This approach is the one taken by \cite{OpenCL}, where thousands of threads are virtually running in parallel, and is described in paragraph 5.

\section{Pixel splitting%
  \label{pixel-splitting}%
}

Pixel splitting is what occurs when a pixel of the detector spans over more than one of the bins of the histogram.
When this happens, the contribution to each of the bins involved is assumed to be
proportional to the area of the pixel segment that falls into that bin.
The goal behind the addition of extra complexity to the code is that the
results obtained this way ought to be less noisy than the case where pixel splitting is ignored.
This becomes more apparent when the number of pixels falling into each bin
is small like for example for 2D integration.
Figure \DUrole{ref}{bidimentional} presents the results of such an integration, performed using histograms
on the top image, i.e. without pixel splitting.
Some high frequency patterns are visible near the beam center on the left-hand side of this figure.
The bottom image was produced using pixel splitting and is
unharmed by such defects, which are related to low statistics.
Note that for 2D integration, this transformation looks like an interpolation,
but interpolation neither guarantees the conservation of the signal $\sum{image} = \sum{ weighted\ histogram }$
nor that of the pixels $\sum{ unweighted\ histogram } = number\ of\  pixels$.\begin{figure}[]\noindent\makebox[\columnwidth][c]{\includegraphics[width=\columnwidth]{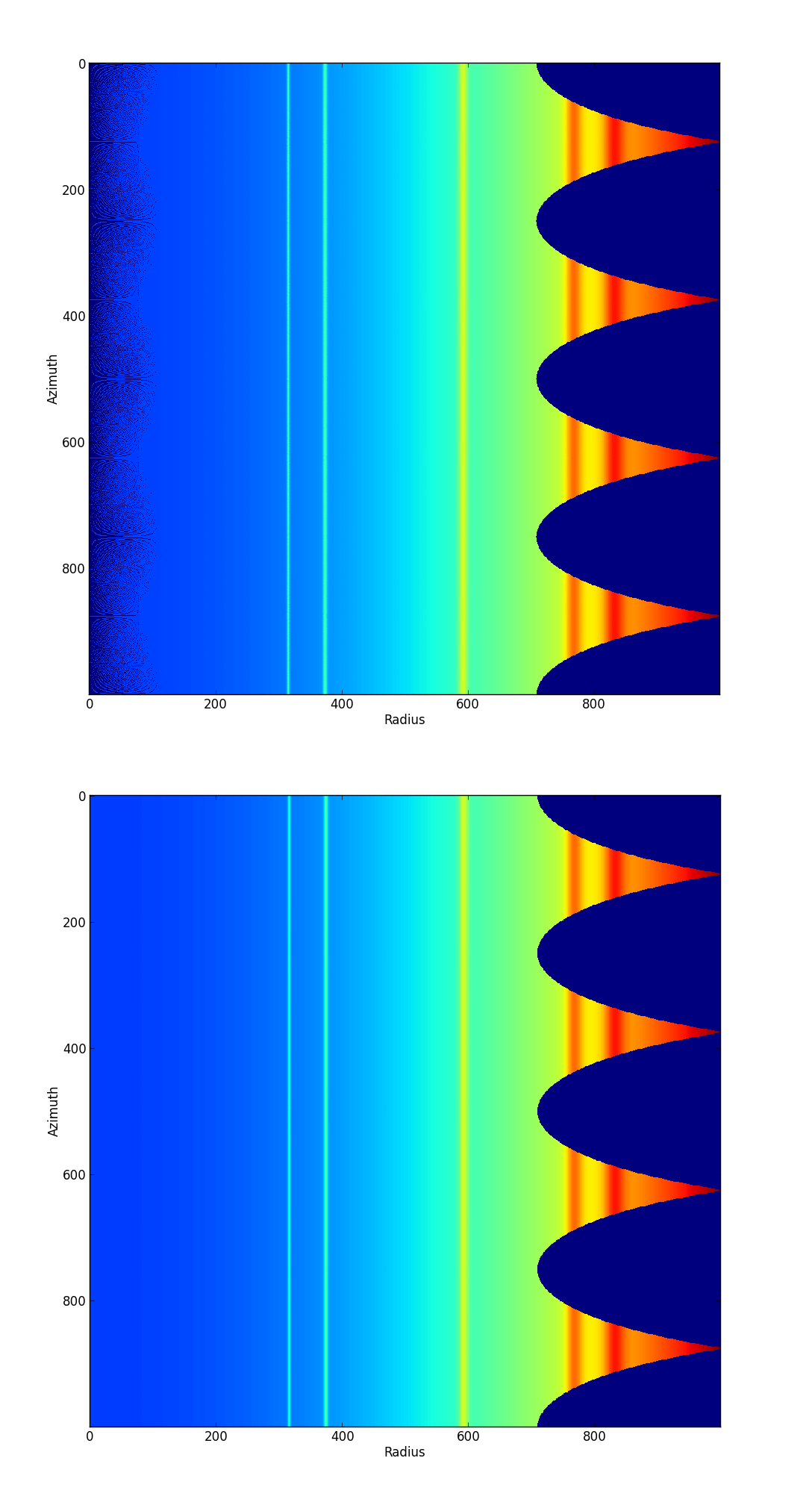}}
\caption{Bi-dimensional azimuthal integration of the gold diffraction image using (bottom) or not (top) pixel splitting \DUrole{label}{bidimentional}}
\end{figure}

\subsection{Bounding Box%
  \label{bounding-box}%
}

The first way pixel splitting was implemented was with a bounding box like in Fit2D \cite{FIT2D}.
In this case we are working with an abstraction of the pixel.
This is represented by a rectangular box circumscribing the actual pixel,
with two sides parallel to the radial axis and the other two of unit length.
Presently, instead of calculating the contribution of each segment of the pixel based on its area, we use the area of the bounding box segment instead.
This greatly simplifies the algorithm's flow, providing good performance.

The algorithm loops over all the pixels of the detector, adding their contributions to the appropriate bins.
When the whole pixel falls into only one bin, there is no pixel splitting and the algorithm proceeds as in the case of the simple histogram.
If the pixel spans over more than one bin, the contribution to the two outermost bins (left and right) is calculated first and added to them.
Then, the remaining contribution is evenly distributed among the “internal” bins (if any).
Finally, the ratio of the two histograms is calculated and returned.

The trade-off of using this simplistic pixel splitting is an overestimation of the pixel size, hence a slight blurring of the signal.

\subsection{Full Pixel Splitting%
  \label{full-pixel-splitting}%
}

In an effort to farther improve the quality of the results of the azimuthal integration,
another pixel-splitting scheme was devised,
in which no abstraction takes place and the pixel splitting
works using the area of the actual pixel segments (assuming they are straight lines).
This introduces some additional complexity to the calculations,
making the process a bit slower.

As before, the algorithm first has to check if pixel splitting occurs.
In the case it does not, the pixel is processed like in the case of the simple histogram.
Otherwise the pixel is split according to the following steps.
Firstly, a function for each of the lines that make up the sides of the pixel being processed is defined
by calculating the slope and the point of intersection.
The area of the pixel is also required.
Next, the algorithm loops over the bins that the pixel spans over and proceeds to
integrate the four functions that were previously defined over the bounds of each bin.
Taking the absolute value of the sum of all these integrals
will yield the area of the pixel segment.
Now, the contributions to the histograms are calculated using these areas.
The difficult part here was the definition of the limits of each of the integrals in a
way that would not hinder the performance by adding many conditionals.

\subsection{Discussion on the statistics%
  \label{discussion-on-the-statistics}%
}

Using either of the two pixel splitting algorithms results in some side effects that the user should be aware of:
The fact that pixels contributing to neighbouring bins in the histogram creates some cross-correlation between those bins,
affecting, this way, the statistics of the results in a potentially unwanted manner \cite{Stat}.

\section{More parallelisation%
  \label{more-parallelisation}%
}

For faster execution, one solution is to use many-core systems, such as
Graphical Processing Units (GPUs) or
accelerators, like the Xeon-Phi from Intel.
Those chips allocate more silicon for computing (arithmetic logic units - ALUs)
and less to branch prediction, memory pre-fetching and cache coherency, in comparison with CPUs.
Our duties as programmers is to write the code that maximises the usage of ALUs
without relying on pre-fetcher and other commodities offered by normal processors.

Typical GPUs have tens (to hundreds) of compute units able to schedule and run
dozens of threads simultaneously (in a Single Instruction Multiple Data way).
OpenCL allows the execution of the same code on processors, graphics cards or accelerators (see table \DUrole{ref}{Devices})
but the memory access pattern is important in order to make the best use of them.
Finally, OpenCL uses just-in-time (JIT) compilation, which looks very much
like Python interpreted code when interfaced with \cite{PyOpenCL}
(thanks to the compilation speed and the caching of the generated binary).\begin{table*}
\setlength{\DUtablewidth}{0.8\linewidth}
\begin{longtable*}[c]{|p{0.220\DUtablewidth}|p{0.128\DUtablewidth}|p{0.128\DUtablewidth}|p{0.107\DUtablewidth}|p{0.107\DUtablewidth}|p{0.148\DUtablewidth}|p{0.128\DUtablewidth}|}
\hline

Vendor / driver & 

Intel & 

AMD & 

AMD & 

Nvidia & 

Nvidia & 

Intel \\
\hline

Model & 

2xE5-2667 & 

2xE5-2667 & 

V7800 & 

Tesla K20 & 

GeForce 750Ti & 

Phi 5110 \\
\hline

Type & 

CPU & 

CPU & 

GPU & 

GPU & 

GPU & 

ACC \\
\hline

Compute Unit & 

12 & 

12 & 

18 & 

13 & 

5 & 

4x69 \\
\hline

Compute Element/CU & 

4:AVX & 

1 & 

80 & 

4x8:Warp & 

4x8:Warp & 

16:AVX512 \\
\hline

Core frequency & 

2900 MHz & 

2900 MHz & 

700 MHz & 

705 MHz & 

1100 MHz & 

1052 \\
\hline

Mem. Bandwidth & 

102 GB/s & 

102 GB/s & 

128 GB/s & 

208 GB/s & 

88 GB/s & 

320 GB/s \\
\hline
\end{longtable*}
\caption{Few OpenCL devices we have tested our code on. \DUrole{label}{Devices}}\end{table*}

\subsection{Parallel algorithms%
  \label{parallel-algorithms}%
}

Parallelisation of complete algorithms consists, most of the time, in their decomposition into parallel blocks.
There are a few identified parallel building blocks like:%
\begin{itemize}

\item 

Map: apply the same function on all elements of a vector
\item 

Scatter: write multiple outputs from a single input, needs atomic operation support
\item 

Gather: write a single output from multiple inputs
\item 

Reduction: single result from a large vector input, like an inner product
\item 

Scan: apply subsequently an operation to all preceding elements on an vector like np.cumsum
\item 

Sort: There are optimised sorter for parallel implementation.
\end{itemize}

These parallel building blocks will typically be one individual
kernel or a few, since kernel execution synchronises the global memory in OpenCL.
Parallel algorithmics deal with how to assemble those blocks to implement the required features.

\subsection{Parallel azimuthal integration%
  \label{parallel-azimuthal-integration}%
}

Azimuthal integration, like histogramming, is a scatter operation, and hence requires
the support of atomic operations (in our case with double precision floats).
As Cython does not (yet) support atomic operations, enabling OpenMP parallelisation
results in a module that, while being functional, gives the wrong results (we measured 2\%
errors on 8 cores)

To overcome this limitation, instead of looking at where input pixels go to
in the output curve,
we focus on where the output bin comes from in the input image.
This transformation is called a “scatter to gather” transformation and requires atomic operations.
In our case, it was implemented as a single threaded \cite{Cython} module.

The correspondence between pixels and output bins can be stored in a look-up table (LUT)
together with the pixel weight (ratio of areas) making the integration look like a simple
(if large and sparse) matrix vector product.
The LUT size depends on whether pixels are split over multiple bins
and in order to exploit the sparse structure, both the index and the weight of each pixel have to be stored.

By making this change we switched from a “linear read / random write” forward algorithm to a
“random read / linear write” backward algorithm which is more suitable for parallelisation.
For optimal memory access patterns, the array of the LUT may be transposed depending on the underlying hardware (CPU vs GPU).

\subsubsection{Optimisation of the sparse matrix multiplication%
  \label{optimisation-of-the-sparse-matrix-multiplication}%
}

The compressed sparse row (CSR) sparse matrix format was introduced to
reduce the size of the data stored in the LUT.
This algorithm was implemented both in \cite{Cython}-\cite{OpenMP} and \cite{OpenCL}.
Our CSR representation contains \emph{data}, \emph{indices} and \emph{indptr} (row index pointer) so it is fully
compatible with the \emph{scipy.sparse.csr.csr\_matrix} constructor from \cite{SciPy}.
This representation is a \emph{struct of array} which is better suited to GPUs
(strided memory access) while LUT is an \emph{array of struct}, known to be
better adapted to CPU (better use of cache and pre-fetching).
The CSR approach presents a double benefit: first, it reduces the
size of the storage needed, as compared to the LUT, by a factor two to three,
and gives the opportunity of working with larger images on the same hardware.
Secondly, the CSR implementation in \cite{OpenCL} is using an algorithm based
on multiple parallel reductions
where all threads within a workgroup are collaborating to calculate the
content of a single bin.
This makes it very well suited to run on many-core systems where hundreds
to thousands of simultaneous threads are available.

\subsubsection{About numerical precision%
  \label{about-numerical-precision}%
}

Knowing the tight energy constraints, the future of high performance computing
depends on the capability of programs to use the suitable precision for their calculations.
As our detectors provide a sensitivity of 12 to 20 bits/pixel, performing all calculations
in double precision (with 52 bits mantissa) might seem excessive, the 24 bits mantissa
of single precision float being a better choice for the task (with no precision drop).
Moreover, GPU devices provide much more computing power in single precision than in double.
This factor varies from 2 on high-end professional GPUs like Nvidia Tesla to 24 on most consumer grade devices.

When using \cite{OpenCL} for GPUs we used compensated arithmetics (or \cite{Kahan} summation), to
reduce the error accumulation in the histogram summation (at the cost of more operations).
This allows numerically accurate results to be obtained even on cheap consumer grade hardware with the use of
single precision floating point arithmetic (32 bits).
Double precision operations are currently limited to high-price / high-performance GPUs, optimised exactly for that purpose.
The additional cost of Kahan summation (4x more arithmetic operations) is hidden by smaller data types,
a higher number of single precision units and the fact that GPUs are usually limited by the memory bandwidth anyway.

The performances of the parallel azimuthal integration can reach 750 MPixel/s
on recent computers with a mid-range graphics card.
On multi-socket servers featuring high-end GPUs like Tesla cards, the performances are equivalent, but with the
added benefit of working with multiple detectors simultaneously.

\section{Benchmarks%
  \label{benchmarks}%
}

We present the results from several benchmark tests done using the different algorithm options available in PyFAI.
All benchmarks were performed using the same bounding box pixel splitting scheme and the resulting integrated profiles are of equivalent quality.
Execution speed has been measured using the \emph{timeit} module, averaged over 10 iterations (best of 3).
The processing is performed on 1, 2, 4, 6, 12 and 16 Mpixel images, with pixel ranges of either 16 or 32 bits (int or uint), taken from actual diffraction experiments, which are part of the pyFAI test suite.

One small note on the benchmarks that follow. The casting for the 12 Mpixel image was done by one thread on the CPU.
That is why the processing time of the 16 Mpixel image appears to be shorter than that of the 12 Mpixel one.

\subsection{Choice of the algorithm%
  \label{choice-of-the-algorithm}%
}

The LUT contains pairs of an index and a coefficient, hence it is an \emph{array of struct} pattern which is known to make best use of CPU caches.
On the contrary, the CSR sparse matrix representation is a \emph{struct of array} which is better adapted to GPU.
As we can see in figure \DUrole{ref}{serial-lut-csr}, both LUT and CSR outperform the serial code, and both behave similarly:
the penalty of the \emph{array of struct} in CSR is counter-balanced by the smaller chunk of data to be transferred from central memory to CPU.\begin{figure}[]\noindent\makebox[\columnwidth][c]{\includegraphics[width=\columnwidth]{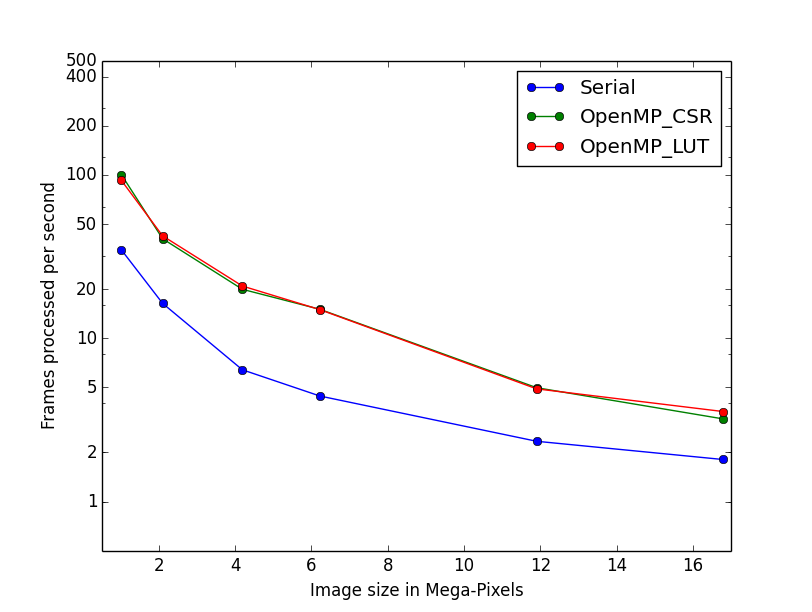}}
\caption{Comparison of azimuthal integration speed obtained using serial implementation versus
parallel implementations with LUT and CSR sparse matrix representation on two Intel Xeon E2667. \DUrole{label}{serial-lut-csr}}
\end{figure}

\subsection{OpenMP vs OpenCL%
  \label{openmp-vs-opencl}%
}

The gain in portability obtained by the use of OpenCL does not mean a sacrifice in performance when the code is run on a CPU,
as we can see in figure \DUrole{ref}{openmp-opencl-intel-amda}: the OpenCL implementation outperforms the OpenMP one, in all the different CPUs it was tested on.
This could be linked to the better use of SIMD vector units by OpenCL.
The dual Xeon E5520 (a computer from 2009), running at only 2.2 GHz shows pretty good performances compared to more recent computers when using OpenMP:
it was the only one with activated hyper-threading.\begin{figure}[]\noindent\makebox[\columnwidth][c]{\includegraphics[width=\columnwidth]{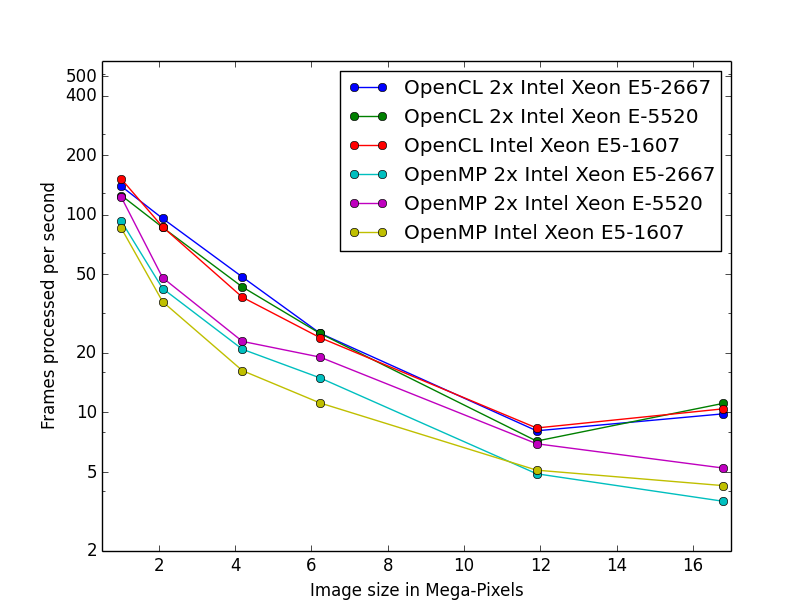}}
\caption{Comparison of the azimuthal integration speed between the OpenMP and OpenCL implementations. \DUrole{label}{openmp-opencl-intel-amda}}
\end{figure}

The choice of the OpenCL driver on CPU affects the performance of PyFAI (figure \DUrole{ref}{openmp-opencl-intel-amdb}):
on the Intel Xeon E5-1607 (Ivy bridge core), the Intel driver clearly outperforms the AMD driver.
This can be attributed to new SIMD instructions (AVX), supported by the Intel driver but not by the AMD one.
On the older Intel Xeon E-5520 (Nehalem core) which lacks those extensions, the difference in speed is much less.\begin{figure}[]\noindent\makebox[\columnwidth][c]{\includegraphics[width=\columnwidth]{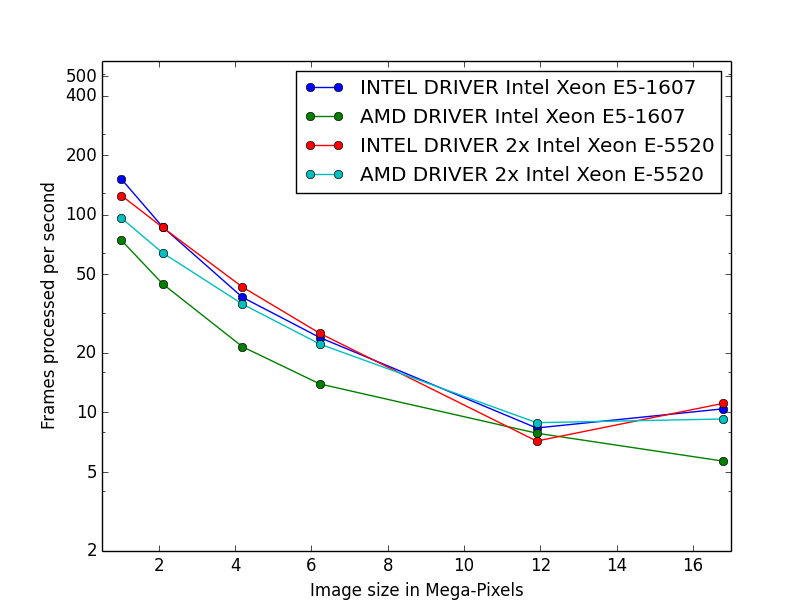}}
\caption{The effects of OpenCL driver selection on performance on different generations of CPUs. \DUrole{label}{openmp-opencl-intel-amdb}}
\end{figure}

\subsection{GPUs and Xeon Phi%
  \label{gpus-and-xeon-phi}%
}

Figure \DUrole{ref}{gpusa} compares the integration speed of the LUT and CSR implementation on two GPUs.
The CSR implementation, thanks to the multiple collaborative parallel reductions, runs much faster on all the devices used, compared to the LUT one.
Another benefit of the CSR implementation when it comes to GPUs is its lower memory usage.
The ATI GPU used in this study features only 1 GB of memory usable by OpenCL, limiting the processable size of the system.
This is the reason the benchmarks stop before reaching the largest image size.
4 Mpixel images are the largest images processable with the LUT implementation, but 12 Mpixel images are processable using the CSR one.\begin{figure}[]\noindent\makebox[\columnwidth][c]{\includegraphics[width=\columnwidth]{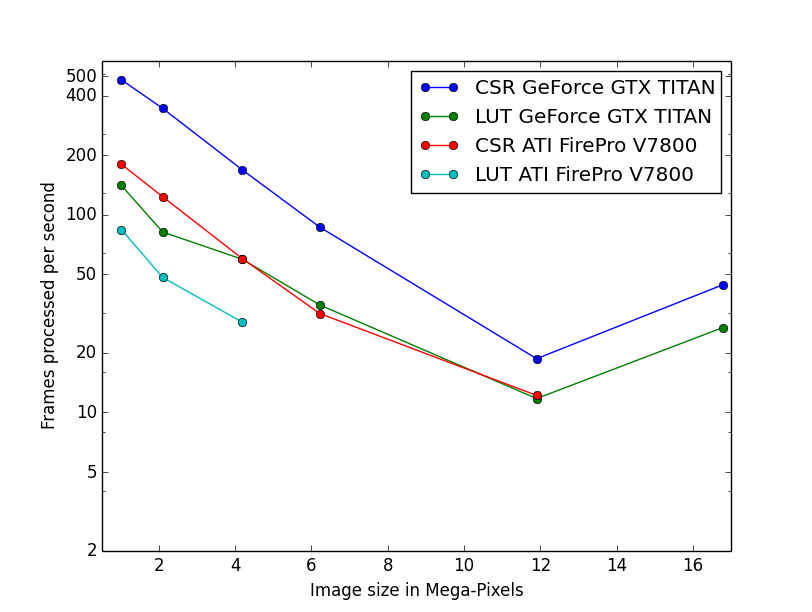}}
\caption{Comparison of the azimuthal integration speed between the LUT and CSR implementations on GPUs. \DUrole{label}{gpusa}}
\end{figure}

In figure \DUrole{ref}{gpusb}, we have gathered the results from all of the many-core devices available to us, including several GPUs as well as Intel's Xeon Phi.
As one can see, Xeon Phi (from 2012) matches the performance of the AMD GPU from 2010.
What is surprising though, is how well the consumer grade Nvidia GeForce 750Ti performs in comparison to high-end \emph{Kepler} cards (Titan, Tesla K20) costing only a fraction of their price.\begin{figure}[]\noindent\makebox[\columnwidth][c]{\includegraphics[width=\columnwidth]{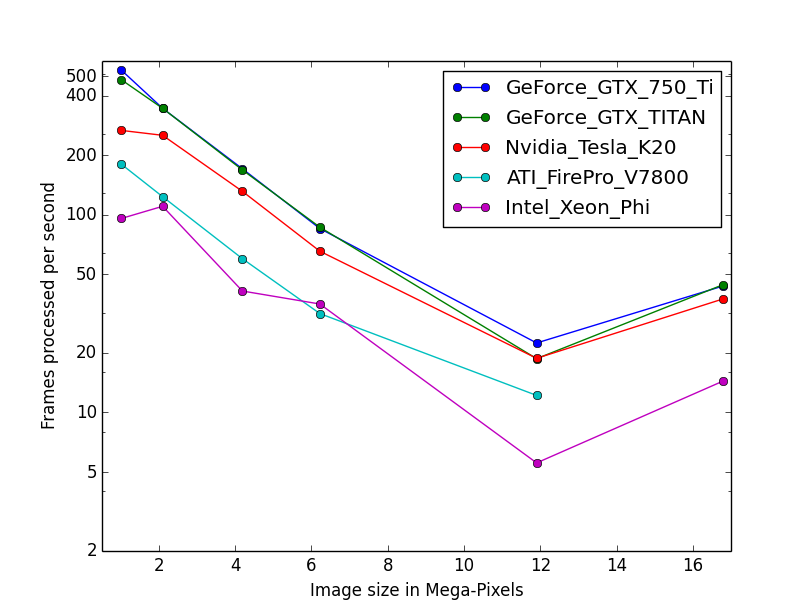}}
\caption{Comparison of the performances for several many-core accelerators: GPUs and Xeon Phi. \DUrole{label}{gpusb}}
\end{figure}

\subsection{Kernel timings%
  \label{kernel-timings}%
}

As stated previously, the benchmark tests were performed using the \emph{timeit} module from Python
on the last line of the code snippet described in section \DUrole{ref}{use}.
One may wonder what is the actual time spent in which part of the OpenCL code and how much is the Python overhead.
This analysis has been done using the profiling tools of OpenCL which measured the execution of every action put in queue.
To be able to perform the azimuthal integration, the image is first transfered to the device (GPU), then casted from integer to float.
All pixel-wise correction (dark current subtraction, flat field normalization, solid-angle and polarization factor correction) are applied in a single pass over each pixel of the image.
Output arrays are initialised to zero, by a separate kernel (memset) before the actual sparse-matrix-dense-vector multiplication.
Finally the three output buffers are retrieved from the device.

Table \DUrole{ref}{profile} shows the execution time measured on the GeForce Titan (controlled by a pair of Xeon 5520).
The first entry in the table is the total execution time at the Python level, as measured by \emph{timeit}: 2 ms,
while the second is the sum of all of the execution times measured by the OpenCL profiler: 1.4 ms, which highlights how little the Python overhead can be (<40\%).
The most time-consuming part of the whole process is by far the memory transfer of the image (H->D meaning Host to Device, 0.8ms).
All vendors are currently working on an unified memory space, which will be available for OpenCL 2.0, which will reduce the time spent in transfers and simplify programming.
Finally the azimuthal integration takes up only 0.4 ms, that is, 20\% of the total run time.
If one focuses only on the timing of the integration kernel, then he would wrongly conclude that pyFAI is able to match the speed of the fastest detectors.
For example, the 2 ms of processing time for a 1 Mpixel image of 32 bit integers, correspond to a processing rate of 2 GB/s, while our fastest storage solutions (solid-state drives)
are currently only able to provide half of that.\begin{table}
\setlength{\DUtablewidth}{0.8\linewidth}
\begin{longtable*}[c]{|p{0.214\DUtablewidth}|p{0.121\DUtablewidth}|}
\hline

Python  total & 

2.030ms \\
\hline

OpenCL & 

1.445ms \\
\hline

H->D image & 

0.762ms \\
\hline

cast & 

0.108ms \\
\hline

memset & 

0.009ms \\
\hline

correction & 

0.170ms \\
\hline

integrate & 

0.384ms \\
\hline

D->H  ratio & 

0.004ms \\
\hline

D->H u\_hist & 

0.004ms \\
\hline

D->H w\_hist & 

0.004ms \\
\hline
\end{longtable*}
\caption{OpenCL  profiling of the integration of a Pilatus 1M image (981x1043 pixels of signed 32 bits integers) on a GeForce Titan, running on a dual Xeon 5520. \DUrole{label}{profile}}\end{table}

\subsection{Configuration and Drivers used%
  \label{configuration-and-drivers-used}%
}

The computer hosting the two Intel Xeon E5-2667 (2x6 cores each, 2.9 GHZ, without hyper-threading, 8x8 GB of RAM) is a Dell PowerEdge R720 with both a Tesla K20 and an Intel Xeon phi accelerator, running Debian 7.
The computer hosting the two Intel Xeon E5520 (2x4cores, 2.27 GHz, hyper-threaded, 6x2 GB of RAM) is a Dell T7500 workstation with two Nvidia GPUs: GeForce 750Ti and Titan, running Debian 7.
The computer hosting the Intel Xeon E5-1607 (1x4cores, 3.0 GHz, without hyper-threading, 2x4 GB of RAM) is a Dell T3610 workstation with two GPUs: Nvidia GeForce 750Ti and AMD FirePro V7800, running Debian 8/Jessie.

In addition to the Debian operating system, specific OpenCL drivers were installed:%
\begin{itemize}

\item 

Intel OpenCL drivers V4.4.0-117 + MPSS stack v3.2.3
\item 

AMD APP drivers 14.4
\item 

Nvidia CUDA drivers 340.24-2
\end{itemize}

\section{Project description%
  \label{project-description}%
}

PyFAI is open-source software released under the GPL license available on GitHub (\url{https://github.com/kif/pyFAI}).
PyFAI depends on Python v2.6 or v2.7 and \cite{NumPy}.
In order to be able to read images from various X-ray detectors, pyFAI relies on the \cite{FabIO} library.
Optional \cite{OpenCL} acceleration is provided by \cite{PyOpenCL}.
Graphical applications for calibration and integration rely on \cite{matplotlib}, \cite{PyQt} and
SciPy \cite{SciPy} for image processing.
A C compiler is needed to build the \cite{Cython} code from the related sources.
PyFAI is packaged and available in common Linux distributions like Debian and Ubuntu but it is also tested and functional under Windows and MacOSX.

\section{Conclusions%
  \label{conclusions}%
}

This contribution shows how one of the most central algorithm in crystallography has been implemented in Python,
optimised in Cython and ported to many-core architectures using PyOpenCL.
A 15x speed-up factor has been obtained by switching from binary code to the OpenCL code running on GPUs (400x vs NumPy).
Some of the best performances were obtained on a mid-range consumer grade Nvidia GeForce 750Ti thanks to the new \emph{Maxwell} generation chip
running as fast as high-end graphics based on the \emph{Kepler} architecture (like the Titan), and literally outperforming
both AMD GPUs and the Xeon-Phi accelerator card.
Programming CPUs in parallel is as easy as programming GPUs via the use of PyOpenCL interfaced with Python.

\section{Acknowledgements%
  \label{acknowledgements}%
}

Claudio Ferrero (head of the Data Analysis Unit) and Andy Götz (head of the Software Group) are acknowledged for supporting the development of pyFAI.
The porting of pyFAI to OpenCL would have not been possible without the financial support of LinkSCEEM-2 (RI-261600), granting the contracts of
Dimitris Karkoulis who started the GPU porting, Zubair Nawaz who ported image distortion and one of the authors (G. Ashiotis) who is working on CSR, pixel splitting and other algorithms.
Finally, the authors would like to acknowledge their colleagues involved in the development of the library, especially Aurore Deschildre and Frédéric Picca.
The authors would like to thank all X-ray beam-lines promoting pyFAI and providing resources to further develop it: ESRF BM01, ID02, ID11, ID13, ID15, ID16, ID21, ID23, BM26, ID29, BM29 and ID30;
and also in other institutes like Soleil, Petra3, CEA, APS who provide feedback, bug reports and patches to the library.

\end{document}